\documentclass[letterpaper,journal]{IEEEtran}

    \usepackage[pdftex]{graphicx}
    \graphicspath{{../pdf/}{../jpeg/}}
    \DeclareGraphicsExtensions{.pdf,.jpeg,.png}  
    \usepackage[cmex10]{amsmath}
    \usepackage{hyperref}
    \usepackage{algpseudocode}
    \usepackage{array}
    \usepackage{mdwtab}
    \usepackage{eqparbox}
    \usepackage{url}
    \usepackage{comment}
    \usepackage{subfigure}
    \usepackage{romannum}
    \usepackage{amssymb}
   \usepackage{threeparttable} 
   \usepackage{lipsum}

    \usepackage[utf8]{inputenc}
    \usepackage{algorithm}

    \usepackage{booktabs} 
    \usepackage{cite}
\usepackage{caption, multirow, makecell}
\usepackage{siunitx}
\usepackage[version=4]{mhchem}

\usepackage{float}
\usepackage{adjustbox}

\usepackage[T1]{fontenc}

\usepackage{tabularx} 
\IEEEoverridecommandlockouts

\begin{document}
\bstctlcite{IEEEexample:BSTcontrol}

\title{WirelessJEPA: A Multi-Antenna Foundation Model using Spatio-temporal Wireless Latent Predictions}

\author{Viet Chu,~\IEEEmembership{Student~Member,~IEEE}, Omar~Mashaal,~\IEEEmembership{Student~Member,~IEEE}, 
        and Hatem~Abou-Zeid,~\IEEEmembership{Member,~IEEE}%

\thanks{The authors are with the Department of Electrical and Software Engineering,
University of Calgary, Calgary, AB T2N 1N4, Canada
(e-mail: viet.chu, omar.mashaal1, hatem.abouzeid@ucalgary.ca).}
}

 

\maketitle

\begin{abstract}

We propose \emph{WirelessJEPA}, a novel wireless foundation model (WFM) that uses the Joint Embedding Predictive Architecture (JEPA). WirelessJEPA learns general-purpose representations directly from real-world multi-antenna IQ data by predicting latent representations of masked signal regions. This enables multiple diverse downstream tasks without reliance on carefully engineered contrastive augmentations. To adapt JEPA to wireless signals, we introduce a 2D antenna–time representation that reshapes multi-antenna IQ streams into structured grids, allowing convolutional processing with block masking and efficient sparse computation over unmasked patches. Building on this representation, we propose novel spatio-temporal mask geometries that encode inductive biases across antennas and time.
We evaluate WirelessJEPA across six downstream tasks and demonstrate it's robust performance and strong task generalization. Our results establish that JEPA-based learning as a promising direction for building generalizable WFMs.
\end{abstract}

\IEEEoverridecommandlockouts

\IEEEpeerreviewmaketitle


\section{Introduction}

\IEEEPARstart{T}{he} vision of AI-native 6G centers on seamlessly integrating intelligence across the network stack. While deep learning has delivered promising results in wireless tasks such as modulation recognition, channel estimation, and sensing, most of these advances are driven by supervised learning (SL). Despite these successes, SL-based approaches suffer from fundamental limitations: acquiring large labeled data is costly and models typically do not generalize across tasks.

Wireless Foundation Models (WFMs) have recently emerged to mitigate these limitations by learning general-purpose representations that can be efficiently adapted to diverse tasks. This direction is motivated by recent advances in self-supervised learning  (SSL) advances in vision and language domains showing that transferable, multi-task representations can be learned from \emph{unlabeled} data.
In the wireless domain, prior work has explored WFMs across a range of input modalities, from structured representations such as channel state information (CSI) and spectrograms~\cite{r_other_7, palhares2025csi2vec, r_other_6} to raw IQ streams, including multi-task SSL approaches~\cite{Kanu_IQ}, and our prior work, IQFM~\cite{mashaal2025iqfm}, which served as an early step toward scalable foundation models operating directly on raw IQ.
Other related works on SSL for raw IQ signals have also been proposed \cite{cheraghinia2025}; however, these approaches primarily focus on in-distribution generalization within a single data modality or task. As a result, they do not support the diverse, out-of-distribution downstream tasks typically associated with foundation models, nor do they consider multi-antenna IQ streams, which are central to this letter.

\begin{figure*}[!t]
\centering
\includegraphics[width=0.75\linewidth,
                 height=1.40in,
                 trim={0.1cm 5.5cm 2cm 1.4cm}, 
                 clip]{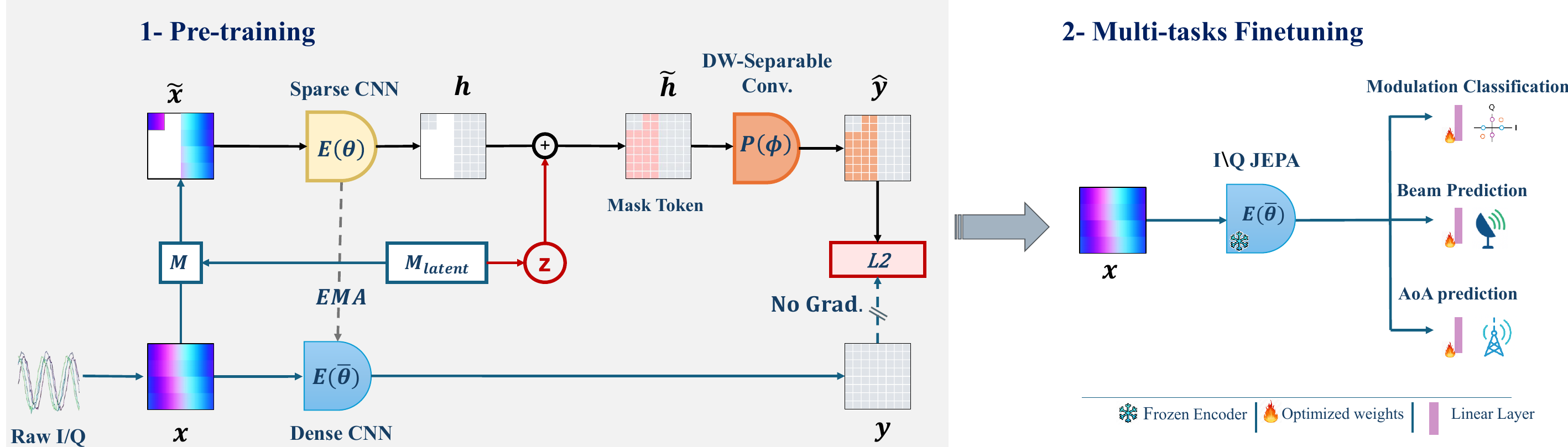}
\caption{ Proposed Wireless-JEPA MIMO Foundation Model.}
\label{fig:iqjepa_arch}
\end{figure*}

While IQFM ~\cite{mashaal2025iqfm} enabled diverse wireless tasks, it relied on contrastive SSL with carefully crafted augmentations to effectively learn spatial and temporal relationships of the multi-antenna IQ data. Poorly designed augmentations were found to encourage shortcut features that fail to generalize beyond the training distribution. More broadly, learning representations by discriminating artificially constructed views may provide a weak inductive bias to model the underlying dynamics of raw signals ~\cite{Li2022Unsupervised}. These observations motivate moving beyond contrastive objectives toward prediction-based, non-contrastive approaches - such as Joint-Embedding Predictive Architectures (JEPA) \cite{assran2023ijepa}. 
JEPA-based models avoid reliance on handcrafted augmentations and instead learn by predicting \emph{latent representations} of masked input regions. They have demonstrated strong effectiveness in multi-task vision and robotics applications by encouraging the learning of transferable representations from unlabeled data. These properties motivate their adoption for wireless multi-antenna IQ representation learning, which we introduce in this letter. 
Our main contributions are:
\vspace*{-1mm}
\begin{itemize}
    \item  
    We propose \textbf{WirelessJEPA}, the first JEPA-based wireless foundation model. WirelessJEPA enables multiple diverse tasks with a single model by learning to predict the wireless latent representations of a real-world multi-antenna IQ dataset. This is accomplished without requiring carefully crafted augmentations as in contrastive learning. 
    \item 
    \textbf{2D antenna-time grids}. We demonstrate how to apply a CNN-JEPA framework \cite{kalapos2024cnnjepa} to wireless data by first shaping multi-antenna IQ streams into a 2D antenna–time grid and then applying convolutions directly on it. This architectural step enables the use of block masking and a computationally efficient sparse-convolution mechanism that operates only on unmasked patches.  
    \item 
    \textbf{Novel spatio-temporal mask geometries.} We hypothesize that \emph{\textbf{structured}} spatio-temporal masking will be beneficial for wireless data over random masking typically applied in JEPA. We therefore propose and thoroughly analyze the benefits of three mask geometries that introduce spatio-temporal inductive biases that encourage the model to learn dependencies across both antennas and time without relying on handcrafted augmentations.
    \item 
    \textbf{Robust performance.} 
     We provide a comprehensive analysis of WirelessJEPA's performance across 6 diverse downstream tasks: angle-of-arrival estimation, modulation classification, RF fingerprinting, protocol classification, GNSS jamming, and interference classification. Our results demonstrates the effectiveness of such JEPA-based frameworks to build WFMs and provides a foundation for further research toward generalizable WFMs using JEPA. 
\end{itemize}

\section{Problem Formulation}
\label{sec:problem-formulation}
\subsection{Signal Model}
We consider a URA receiver observing raw complex baseband signals. Each sample $x$ is represented as a real-valued tensor:

$$ x \in \mathbb{R}^{C \times H \times W} $$
where $C = 2$ corresponds to the in-phase (I) and quadrature (Q) components, $H$ is the number of receiver antennas, and $W$ is the number of time samples. Following \cite{mashaal2025iqfm}, we apply unit-max normalization $x \leftarrow x/\max(|x|)$. The receiver USRPs share an OctoClock reference, while the transmitter is unsynchronized; thus residual CFO/timing drift is not software-compensated. Fig.~\ref{fig:iqjepa_arch} visualizes $x$.

\subsection{Learning Objective}
Given a large, unlabeled dataset of raw IQ signals, $\mathcal{D} = \{x_1, x_2, \dots, x_N\}$, our objective is to learn a powerful representation model, $E(\theta)$, in a self-supervised manner.
The goal is for this single, pre-trained encoder to produce representations $h = E(\theta, x)$ that are transferable, and capture the fundamental properties of the wireless signals. These representations should then be highly effective for a wide variety of downstream wireless tasks with low-shot and lightweight adaptation.Concretely, we optimize $\theta$ (and predictor parameters $\phi$) by minimizing the JEPA feature-prediction loss over masked regions (Sec.~III, Alg.~1).

\section{WirelessJEPA}
\label{sec:JEPA}
This section presents the WirelessJEPA methodology, including the architecture and masking strategies, training algorithms, and datasets for pretraining and downstream tasks.
\subsection{Proposed Architecture: Learning from the Latent Space} 
Instead of learning by masking and reconstructing patches of \emph{raw} multiple antenna IQ streams, WirelessJEPA operates in the \emph{latent} space. As shown in Fig.~\ref {fig:iqjepa_arch}, the model learns by masking spatial–temporal regions of the input IQ samples $\tilde{x}$ and then predicting their latent representations $\hat{y}$. This is accomplished in two stages, the first being to learn an encoded representation of the \emph{unmasked} multi-antenna IQ region using an encoder $E(\theta)$. These representations are then used by a predictor $P(\phi)$ to predict the latent representations of the \emph{masked} region. After training is complete, the learned encoder $E(\theta)$ is used as the WFM for multiple downstream tasks.

In more detail, we adopt CNN-JEPA\cite{kalapos2024cnnjepa}, a JEPA variant specifically designed for convolution architectures, which is composed of three main components: a sparse convolution context encoder $E(\theta)$, a lightweight predictor $P(\phi)$, and a momentum teacher encoder $E(\bar{\theta})$. The context encoder extracts meaningful representations from unmasked IQ signals. The lightweight predictor enables latent-space prediction while ensuring that semantic information is learned by the encoder itself. The momentum teacher encoder operates on the full IQ representation and generates stable target embeddings for the masked regions. The teacher’s parameters are not directly trained but are instead updated as an exponential moving average of the student’s parameters $E(\theta)$, providing a stable learning target.

\textbf{Encoder Architecture.} We use ShuffleNetV2-x0.5 for its compact efficiency and to match the IQFM baseline for a controlled comparison between contrastive learning and JEPA \cite{mashaal2025iqfm}. The 2-channel I/Q input mapping is described in \ref{sec:masking}

\textbf{Pretraining Dataset.}
WirelessJEPA is pre-trained on the same over-the-air multichannel IQ dataset as IQFM \cite{mashaal2025iqfm}, collected with a MIMO testbed using USRP X300 receivers. It includes 7 waveform types and 225 AoA classes, captured at 1 and 10~MSps under real multipath. Each recording is segmented into $x\in\mathbb{R}^{2\times4\times256}$ (4 antennas, 256 time samples).

\subsection{Proposed Spatio-temporal Masking Strategies} \label{sec:masking}
A key challenge in applying CNN-based SSL to wireless signals lies in the input signal's geometry. As defined in Section \ref{sec:problem-formulation}, our input tensor $x \in \mathbb{R}^{C \times H \times W}$ has a non-square aspect ratio ($H=4$ antennas and $W=256$ time samples). This is incompatible with the standard multi-block masking strategy of CNN-JEPA \cite{kalapos2024cnnjepa}, which assumes a more uniform 2D layout. To address this, we first apply an antenna-dimension upsampling step via nearest-neighbor interpolation. Concretely, for an input $x_{raw} \in \mathbb{R}^{2 \times 4 \times 256}$, each of the $H=4$ antenna rows is up-sampled 64 times, such that
$$x_{c,i,t} = (x_{raw})_{c,\,\lfloor i / 64 \rfloor,\,t}, \quad c \in \{\mathrm{I}, \mathrm{Q}\},\ i,t \in [0, 255],$$
producing a uniform tensor $x \in \mathbb{R}^{2 \times 256 \times 256}$. This transformation creates a square antenna–time plane, enabling the direct application of 2D masking techniques discussed next.

\begin{figure}[t]
  \centering
  \subfigure[Orig.\label{fig:mask_orig}]{
    \includegraphics[width=0.067\textwidth]{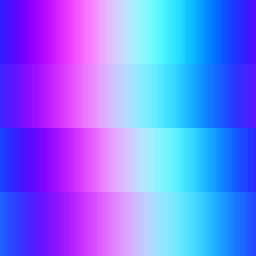}}
  \hfill
  \subfigure[Random\label{fig:mask_random}]{
    \includegraphics[width=0.067\textwidth]{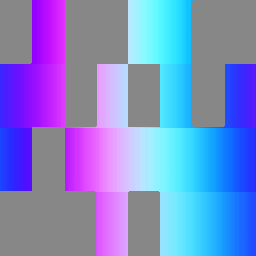}}
  \hfill
  \subfigure[Antenna\label{fig:mask_antenna}]{
    \includegraphics[width=0.067\textwidth]{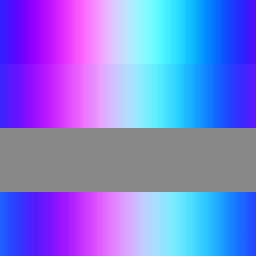}}
  \hfill
  \subfigure[Time\label{fig:mask_time}]{
    \includegraphics[width=0.067\textwidth]{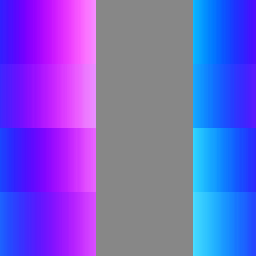}}
 \hfill
\subfigure[M.block\label{fig:mask_multiblock}]{
    \includegraphics[width=0.067\textwidth]{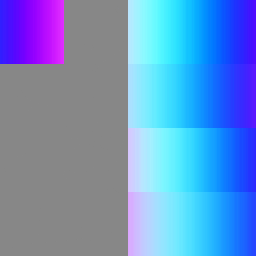}}
\caption{Proposed multi-antenna IQ masking strategies.}
  \label{fig:mask_viz}
\end{figure}

\textbf{Mask Geometry.} Unlike CNN-JEPA which employs random masking, we propose \emph{structured} spatio-temporal masking  that we expect will be superior to random masking. 
Figure~\ref{fig:mask_viz} visualizes the three proposed masking strategies (antenna, time and multi-block) and their resulting occlusion geometries.
These spatio-temporal masking strategies inject an inductive bias during the masking process as follows: antenna masking encourages learning cross-antenna embeddings that enable predicting latent representations of the masked antenna, while time masking emphasizing predicting temporal latent signal representations. Multi-block masking introduces a simple, partial spatio-temporal structure by grouping occlusions into contiguous regions and encourages both cross-antenna and temporal learning. We thoroughly investigate and verify these trends in the experiment results.

\subsection{Training Algorithm}
The overall training procedure is detailed  Algorithm~\ref{alg:iqjepa-train} and consists of two main stages: (i) a context-encoder forward path that extracts latent embeddings and (ii) a latent-space prediction stage that extracts \emph{masked-region} IQ embeddings.

\textbf{The Encoder and Mask Token.} The context encoder $E(\theta)$ processes the masked input $\tilde{x}$, producing the feature maps
\[
    h = E(\theta, \tilde{x}, M_{\text{latent}}).
\]
 
However, since standard convolutions propagate information across all receptive fields, including masked regions, a sparse convolution process is applied to regulate data flow. As illustrated in Algorithm~\ref{alg:sparse-encoder-proc}, rather than enforcing sparsity within the network architecture, an upsampled version of $M_{\text{latent}}$ is reapplied after each layer (convolution or pooling) to suppress activations originating from masked regions. This masking propagation control is adopted for simplicity of implementation, following the CNN-JEPA \cite{kalapos2024cnnjepa} design. Lastly, the learnable mask token $z$ is then inserted at masked indices of the latent space, resulting in the final context features $\tilde{h}$.

\begin{algorithm}[t]
\caption{Proposed WirelessJEPA Training Algorithm}
\label{alg:iqjepa-train}
\begin{algorithmic}[1]

\State \textbf{Input:} Mask tensor $M_{\text{latent}}$; multi-antenna IQ sample $x_{raw}$ 
\State \textbf{Initialize:} Encoder $E(\theta)$, predictor $P(\phi)$, teacher $E(\bar{\theta})$
\State \textbf{Output:} Final trained encoder $E(\theta)$
\Procedure{WirelessJEPA\_Train}{$x_{raw}$, $M_{\text{latent}}$}
\Repeat \Comment{for each training iteration}

  \State Upsample $x_{raw}$ along antenna dimension to derive:  
  \State  $x \in \mathbb{R}^{2 \times 256 \times 256}$
  \State Upsample latent mask $M_{\text{latent}}$ to derive mask $M$
  \State Construct masked input $\tilde{x} = x \odot M$ 
  \State $h = E(\theta, \tilde{x}, M_{\text{latent}} )$ \\ \Comment{\textbf{WirelessJEPA\_Forward}, Alg.~\ref{alg:sparse-encoder-proc}}
  \State Insert mask token $z$ at masked latent indices:  
  \State $\tilde{h} = h \oplus z$
  \State $\hat{y} = P(\phi, \tilde{h})$ \Comment{predict masked latent targets}
  \State $y = E(\bar{\theta}, x)$ \Comment{teacher encoder (dense forward)}
  \State Compute regression loss over masked indices $\mathcal{M}$:
  \State $\mathcal{L} = \frac{1}{|\mathcal{M}|} \sum_{(i,j)\in\mathcal{M}} \|\hat{y}_{i,j} - y_{i,j}\|_2^2$
  \State Update $\theta,\phi$ by gradient descent on $\mathcal{L}$
  \State Update momentum coefficient $\tau$
  \State Update teacher $E(\bar{\theta})$ via EMA: $\bar{\theta} \leftarrow \tau \bar{\theta} + (1-\tau)\theta$
\Until{convergence}
\State \Return $E(\theta)$ \Comment{trained encoder}
\EndProcedure
\end{algorithmic}
\end{algorithm}

\begin{algorithm}[t]
\caption{WirelessJEPA Forward Pass (Sparse Conv.)}
\label{alg:sparse-encoder-proc}
\begin{algorithmic}[1]

\State \textbf{Input:} Mask tensor $M_{\text{latent}}$; masked IQ tensor $\tilde{x}$
\State \textbf{Output:} latent embeddings $h$ 
\Procedure{WirelessJEPA\_Forward}{$E$,$\tilde{x}$,$M_{\text{latent}}$}
\State $h \gets \tilde{x}$ \Comment{initialize hidden state}
  \For{each layer $L$ in encoder $E$}
    \State $M_{\text{layer}} \gets \textbf{AdaptMaskToLayer}(M_{\text{latent}},\ \text{size}(h))$ \\
    \Comment{adjust mask size to layer dimensions}
    \If{$L$ is BatchNorm or LayerNorm} \\
        \Comment{normalize only unmasked positions}
        \State $h_{\text{active}} \gets L(h_{\text{active}})$ 
        \State $h_{\text{inactive}} \gets 0$
    \Else \Comment{for Conv, Pool, or other layer types}
        \State $h \gets L(h)$
        \State $h \gets h \odot M_{\text{layer}}$ \Comment{keep masked locations zero}
    \EndIf
  \EndFor

  \State \Return $h$ \Comment{latent features with masked regions}
\EndProcedure
\end{algorithmic}
\end{algorithm}

\textbf{Predictor and Teacher Models.} The lightweight predictor $P(\phi)$ operates on $\tilde{h}$ to generate latent predictions
\[
    \hat{y} = P(\phi, \tilde{h}),
\]
restricted to masked locations only. To ensure efficiency without compromising accuracy, we implement $P(\phi)$ with depthwise separable convolutions as in CNN-JEPA \cite{kalapos2024cnnjepa}. 

The teacher encoder $E(\bar{\theta})$ shares the same architecture as $E(\theta)$ but operates in a fully dense manner, without any masking or sparsity applied. It processes the unmasked input $x$ to produce stable target embedding representations 
\[ 
    y = E(\bar{\theta}, x)
\]
These targets act as slowly evolving anchors for the predictions produced by the context encoder. Unlike the context encoder, the teacher parameters $\bar{\theta}$ are not updated by gradient descent, but instead follow an exponential moving average (EMA) update as in standard SSL. At each training iteration $t$, the teacher parameters are updated as:
\[ 
    \bar{\theta}_{t} \leftarrow \tau \bar{\theta}_{t-1} + (1 - \tau)\theta_{t},
\]
where $\tau \in [0, 1)$ is a momentum coefficient controlling the update rate.

The model is trained by minimizing an L2 regression loss over masked indices:
\[
    \mathcal{L} 
    = \frac{1}{|\mathcal{M}|} \sum_{(i,j)\in\mathcal{M}} \left\| \hat{y}_{i,j} - y_{i,j} \right\|_2^2,
\]
where $\mathcal{M} = \{ (i,j) \mid M_{i,j} = 0 \}.$ Gradients flow only through $E(\theta)$ and $P(\phi)$, encouraging the model to infer missing spatial--temporal latent representation from the available context rather than directly reconstructing raw waveforms.

\subsection{Downstream Tasks}
To evaluate both in-distribution and out-of-distribution generalization, we assess WirelessJEPA on the two testbed tasks and five additional downstream datasets:
\begin{itemize}
    \item \textbf{Testbed (In-Distribution) \cite{mashaal2025iqfm}:} Modulation classification and AoA estimation using the same pre-training dataset.
    
    \item \textbf{POWDER RF Fingerprinting \cite{reus2020trust}:} 4-device classification from WiFi signals collected across distinct hardware.
    
    \item \textbf{RML2016.10a  \cite{oshea2017deep2}:} Modulation classification with 11 modulation types under varying SNRs ($-20$~dB to $+18$~dB), using $(2,1,128)$ single-channel I/Q inputs.

    \item \textbf{GNSS Jamming (6-Class) \cite{swinney2021raw}:} Raw I/Q classification of six GNSS jamming conditions 
    generated using the Zenodo GNSS jamming synthesis framework.

    \item \textbf{WiFi (4-Class) \cite{belgiovine2024tprime}:} OTA classification of IEEE\,802.11ax/b/n/g protocols using $(2,1024)$ raw I/Q windows extracted from Room~A captures.

  \item \textbf{5G NR Interference:} 
        OTA 5G~NR IQ dataset collected at 3.415~GHz (Band~78) with 10~MHz bandwidth. Four interference classes (LFM, collision, bleedover, wideband) are injected using a USRP~X310 and received by a USRP~B210, under varying gain.
\end{itemize}

\section{Experiments}
\label{sec:experiments}

To evaluate the quality of our pre-trained representations, we adopt a low-shot transfer learning protocol. We freeze the pre-trained encoder and attach lightweight classifiers: linear probing and non-parametric k-nearest neighbor (k-NN) classification. This evaluation is performed on the aforementioned seven downstream tasks. For comparison, we benchmark against the contrastive learning framework of IQFM \cite{mashaal2025iqfm}, which employs an I/Q encoder trained with spatial–temporal augmentations.

\textbf{Experimental Setup.} For consistency with the baseline, we use the same ShuffleNetV2 (0.5×) \cite{ma2018shufflenetv2} encoder as IQFM. Training is performed for 100 epochs using the AdamW 
optimizer with cosine learning rate decay and exponential moving average (EMA) teacher updates, with the momentum $\tau $ gradually increased from 0.996 to 1.0. The predictor comprises three depthwise separable convolutional layers with Batch Normalization and ReLU activation, following CNN-JEPA \cite{kalapos2024cnnjepa}.

\subsection{Mask Geometry and Inductive Bias}

\begin{figure}[b]
    \centering
    \includegraphics[width=0.31\textwidth]{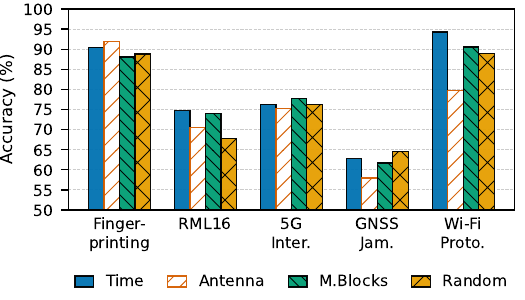}
    \caption{Ablation of WirelessJEPA masking strategies linear probing accuracy at 500-shots on five OOD downstream tasks.}

    \label{fig:ood_ablations}
\end{figure}

We first investigate how mask geometry shapes the inductive bias of WirelessJEPA relative to the multi-antenna I/Q data structure. Table \ref{tab:mask_ablation_combined} and Fig.\ref{fig:ood_ablations} summarize the WirelessJEPA results with different mask geometries for the in-distribution and OOD tasks respectively. The results
indicate that unstructured \emph{random} masking is generally suboptimal for capturing complex signal dependencies. Our intuition is that random masking, which scatters isolated patches, does not effectively leverage the physical continuity of the signal, resulting in an encoder with less expressive representations. In contrast, time masking induces a strong temporal bias, which aligns well with the predominantly temporal nature of several OOD tasks. Accordingly, time masking achieves the highest accuracy on 3 OOD tasks and leads the in-distribution modulation task with 80.75\% 1-shot accuracy. This verifies the temporal inductive bias that enables few-shot modulation classification. However, this temporal focus reduces spatial discrimination, reflected in the lowest 1-shot AoA accuracy (2.70\%).

Conversely, antenna masking enforces a spatial inductive bias by occluding entire antenna inputs. This encourages the model to rely on spatially driven variations such as inter-antenna phase relationships. As a result, antenna masking achieves the highest 1-shot AoA accuracy (40.39\%) but is the least effective on modulation classification tasks. Multi-block masking which introduces a contiguous spatiotemporal structure by grouping occlusions into contiguous regions results in an overall robust performance in both temporal and spatial tasks. These results highlight that enforcing local continuity helps JEPA learn more robust features than completely unstructured occlusions.

\begin{table}[t]
  \centering
  \begin{threeparttable}
    \caption{Impact of WirelessJEPA mask geometry on modulation and AoA linear probe accuracy (\%).}
    \label{tab:mask_ablation_combined}
    
    \renewcommand{\arraystretch}{1.25}
    
    \begin{tabular}{lccccc}
      \toprule
      \multirow{2}{*}{Model} & \multirow{2}{*}{Patch Size} & \multicolumn{2}{c}{1-shot} & \multicolumn{2}{c}{100-shot} \\
      \cmidrule(lr){3-6}
       & & Mod & AoA & Mod & AoA \\
      \midrule
      M.block       & $64 \times 32$   & 73.08 &  5.76 & 99.87 & 93.74 \\
      Random        & $64 \times 32$   & 53.95 & 12.56 & 94.23 & 98.64 \\
      Antenna       & $64 \times 256$  & 48.82 & 40.39 & 99.99 & 99.87 \\
      Time          & $256 \times 32$  & 80.75 &  2.70 & 99.98 & 22.35 \\
      \midrule
      Supervised    & --               & 14.27 &  0.45 & 95.66 & 98.81 \\
      IQFM          & --               & 60.48 & 32.42 & 99.95 & 99.33 \\
      \bottomrule
    \end{tabular}

  \end{threeparttable}
\end{table}

\subsection{Out-of-Distribution Performance: WirelessJEPA vs. IQFM} 

\begin{figure}[t]
    \centering
    \includegraphics[width=0.31\textwidth]{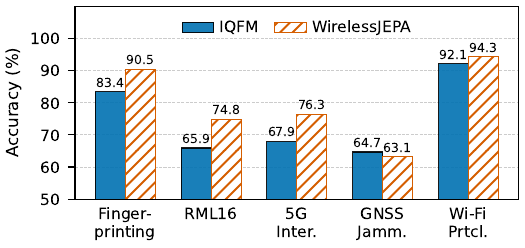}
    \caption{WirelessJEPA vs IQFM linear probing accuracy at 500-shot on unseen OOD downstream tasks.}

    \label{fig:ood_results}
\end{figure}
   
    We evaluate the generalization capability of WirelessJEPA against IQFM, a state-of-the-art contrastive learning framework, across multiple OOD datasets. We selecting the time-masked model based on its strong performance on the predominantly temporal downstream tasks (as established in Section IV-A) - and the results in Fig.~\ref{fig:ood_results} show that WirelessJEPA outperforms the baseline on four out of five of the OOD datasets. The JEPA masked latent prediction approach yields substantial gains, achieving absolute improvements of 8.9\% on RML16 (74.8\% vs.\ 65.9\%) and 8.4\% on 5G Interference (76.3\% vs.\ 67.9\%), while also surpassing IQFM on RF Fingerprinting (90.5\% vs.\ 83.4\%). Although it slightly underperforms on GNSS Jamming (63.1\% vs.\ 64.7\%), the overall trend confirms that predictive masking produces more robust and transferable representations for diverse wireless domains than contrastive instance discrimination.

\subsection{In-Distribution Performance: WirelessJEPA vs. IQFM}

Table~\ref{tab:mask_ablation_combined} presents the results of WirelessJEPA and IQFM for the two in-distribution tasks. The AoA task is generally more challenging than the modulation task and requires spatial inductive biases to perform well. IQFM and WirelessJEPA with antennas masking have a comparable performance - WirelessJEPA has an improved 1-shot performance on AoA and IQFM has the edge on 1-shot modulation classification. WirelessJEPA with time and multi-block masking strategies exceed IQFM's modulation accuracy, but are considerably worse in AoA. In summary, the results of the two approaches are quite comparable for the in-distribution tasks but WirelessJEPA excels in the OOD tasks discussed earlier.

\subsection{Representation Space Analysis: Linear vs. k-NN}

To assess the generalizability of the learned representations independent of task-specific fine-tuning, we use k-NN classification on the frozen time-masked WirelessJEPA encoder outputs.
Table~\ref{tab:jepa_knn_linear_merged} shows high k-NN accuracy on modulation (99.9\%) and OOD tasks such as RF Fingerprinting (87.8\%). This indicates that the model forms discriminative clusters without gradient updates. Performance is also very good with the remaining OOD tasks, while the lower k-NN accuracy on AoA reflects the temporal inductive bias of the encoder which captures waveform structure versus spatial phase differences. While linear probing provides additional refinement, the strong non-parametric performance on multiple tasks confirms that WirelessJEPA already captures highly discriminative features.

\begin{table}[t]
\centering
\caption{Linear vs. k-NN accuracy (\%).}
\label{tab:jepa_knn_linear_merged}
\footnotesize
\setlength{\tabcolsep}{6pt}
\renewcommand{\arraystretch}{1.05}
\begin{tabular}{@{}l c c@{}}
\toprule
Task & Linear & k-NN \\
\midrule
\multicolumn{3}{c}{\textit{In-domain (WirelessJEPA)}} \\
\midrule
Modulation (Mod) & 99.92 & 99.91 \\
Angle of Arrival (AoA) & 40.91 & 17.83 \\
\midrule
\multicolumn{3}{c}{\textit{Out-of-domain (Transfer)}} \\
\midrule
RF Fingerprint & 90.45 & 87.82 \\
RML2016.10a & 74.78 & 68.01 \\
5G Inter. & 76.27 & 64.52 \\
GNSS Jamm. & 63.15 & 48.85 \\
Wi-Fi Prtcl. & 94.26 & 84.89 \\

\bottomrule
\end{tabular}
\end{table}


\section{Conclusion}
\label{sec:conclusion}

We presented WirelessJEPA, a JEPA framework that learns representations from masked latent prediction on raw multi-antenna I/Q signals without relying on hand-crafted augmentations. The model achieves strong performance on OOD tasks surpassing contrastive baselines. Our analysis shows that masking geometry directly shapes the inductive bias of the pretrained encoder, with temporal masks favoring waveform structure and antenna masks emphasizing spatial cues. These findings establish JEPA-style pretraining as a practical direction for wireless foundation models and motivate future work on designing masking strategies that better capture both spatial and temporal structure.

\bibliographystyle{IEEEtran}

\bibliography{IEEEabrv,main}

\end{document}